\documentclass[conference]{IEEEtran}
\IEEEoverridecommandlockouts
\usepackage{amsmath,amssymb,amsfonts}
\usepackage{algorithmic}
\usepackage{algorithm}
\usepackage{array}
\usepackage[caption=false,font=normalsize,labelfont=sf,textfont=sf]{subfig}
\usepackage{textcomp}
\usepackage{stfloats}
\usepackage{url}
\usepackage{verbatim}
\usepackage{graphicx}
\usepackage{cite}
\usepackage{xcolor}
\usepackage[hidelinks]{hyperref}
\usepackage[utf8]{inputenc}
\ifdefined\DeclareUnicodeCharacter
  \DeclareUnicodeCharacter{03BC}{\textmu}
\fi
\makeatletter
\def\ps@headings{%
\let\@oddhead\@empty
\let\@evenhead\@empty
\def\@oddfoot{\@IEEEheaderstyle\hfil}%
\def\@evenfoot{\@IEEEheaderstyle\hfil\hbox{}}%
}
\def\ps@IEEEtitlepagestyle{%
\let\@oddhead\@empty
\let\@evenhead\@empty
\def\@oddfoot{\footnotesize \textbf{979-8-3503-3120-2/24/\$31.00 ~\copyright2024 IEEE} \hfill}%
\let\@evenfoot\@empty
}
\makeatother

\def\BibTeX{{\rm B\kern-.05em{\sc i\kern-.025em b}\kern-.08em
    T\kern-.1667em\lower.7ex\hbox{E}\kern-.125emX}}
\begin{document}

\title{Design and Validation of a Very Low-Power Phasor Measurement Unit
\thanks{This work was supported by the DEVCOM Army Research Laboratory (Cooperative Agreement) under Grant W911NF-17-S-0003.}}

\author{
\IEEEauthorblockN{Zachary J. Lythgoe$^{1}$, Thomas F. Long$^{1}$, Michael J. Buchholz$^{1}$, Anthony R. Livernois$^{1}$, \\ Kebba Kanuteh$^{1}$, David R. Allee$^{1}$, Anamitra Pal$^{1}$, Ian R. Graham$^{2}$, Zachary D. Drummond$^{3}$}

\IEEEauthorblockA{%
\\
\textit{$^{1}$School of Electrical, Computer, and Energy Engineering, Arizona State University, Tempe, USA} \\
\textit{$^{2}$Department of Physics and Astronomy, University of Pennsylvania, Philadelphia, USA} \\
\textit{$^{3}$DEVCOM Army Research Laboratory, Adelphi, USA}
}
}

\maketitle

\IEEEpubidadjcol

\begin{abstract}
Phasor measurement units (PMUs) provide a high-resolution view of the power system at the locations where they are placed. As such, it is desirable to place them in bulk in low voltage distribution circuits.
However, the power consumption of a PMU/micro-PMU is in the order of Watts (W) that results in them requiring an external power supply, which in turn increases the overall cost. This work details the hardware design of a PMU capable of measuring and reporting voltage and current phasors for a single-phase system at an average power consumption of only 30.8 mW -- one to two orders of magnitude lower than existing academic and commercial PMUs.
This enables the proposed PMU to run for two weeks using an 11-Wh battery or indefinitely if paired with an inexpensive solar panel. A test-bench developed in accordance with the 2018 IEC/IEEE 60255-118-1 PMU Standard confirms the accuracy of this PMU. Given its low power consumption, the proposed design is expected to accelerate adoption of PMUs in modern distribution grids.
\end{abstract}


\section{Introduction}
Phasor measurement units (PMUs) 
measure voltage and current phasors (magnitude and phase of the fundamental power system frequency) and frequency and rate of change of frequency (ROCOF) at the locations where they are placed. The measurements are timestamped relative to coordinated universal time (UTC), making them synchronized-phasors or \textit{synchrophasors}, for short. By reporting at faster rates than the power system cycle, PMUs can observe dynamic grid behavior (e.g., oscillations, transients) that supervisory control and data acquisition (SCADA) systems cannot \cite{follum_phasors_2021}. A variety of monitoring, protection, and control algorithms at the transmission level have already been developed using PMU data \cite{phadke_synchronized_2017,kezunovic_application_2014}.

At the distribution level, we find PMUs as well as micro-PMUs ($\mu$PMUs). However, a limitation of existing PMUs/$\mu$PMUs is their need for a dedicated power supply which may prevent them from being placed at desired locations in the \textit{secondary} distribution system (e.g., underneath overhead power lines or on Army microgrids without suitable power supplies). Existing PMUs' high-power processors often consume power in excess of 1 W \cite{thakre_economical_2020,romano_high-performance_2017}. The non-contact, single-phase PMU proposed in \cite{liu_wide-area-measurement_2016} that can be deployed underneath overhead power lines, consumes around 5 W. For continual operation in a typical U.S. East Coast environment, this device will need a 600-Wh car battery and a 100-W solar panel. The lowest-power $\mu$PMU found in the literature is estimated to consume 396 mW \cite{das_development_2016}, and will require a 15-Wh battery and 20-W solar panel for its sustained operation. To the best of the authors' knowledge, no previously published work has targeted minimizing power consumption of PMUs (in addition to maintaining reasonable accuracy) as their objective.

\IEEEpubidadjcol
This paper presents a PMU that measures voltage and current phasors, frequency, and ROCOF using non-contact sensors at a reporting rate of 60 frames per second (fps). The proposed PMU consumes an average power of only 30.8 mW, providing a battery lifetime of two weeks using an 11-Wh battery, or a potentially infinite lifetime with sufficient energy harvesting (e.g., an inexpensive 300-mW solar panel in a generally sunny environment). 
This low power consumption also enables alternative energy harvesting techniques such as mounting on a distribution line and inductively coupling with the current it carries, similar to the point-on-wave recorder in \cite{patterson_inductively_2021}. Components for this PMU cost \$140 per device for a batch of five devices, which is comparable to prices of other PMUs.
It is also compliant with the total vector error (TVE) limits of the 2018 IEC/IEEE 60255-118-1 Standard \cite{noauthor_ieeeiec_2018}. Given the exceptionally low power consumption of this sensor, it is capable of inexpensive operation in environments without access to secondary distribution-level power supplies.

\section{Phasor Estimation Algorithm}
\label{Phasor Estimation Algorithms}
This work uses a variation of the measurement class (M-class) reference phasor estimation algorithm provided in Annex D of \cite{noauthor_ieeeiec_2018}. This algorithm was selected as it reduced the digital signal processing power consumption and accommodated the low-power processor on the PMU. The reference phasor estimation algorithm demodulates the sampled data by the nominal frequency (50 Hz or 60 Hz) and then applies a sinc low-pass finite-impulse response (FIR) filter.

The following modifications were made to improve the phasor estimation accuracy over the original algorithm. To increase robustness to noise, the sample rate was quadrupled from 960 Hz to 3840 Hz. The M-class algorithm filter order was extended to 760, and the standard-specified ``reference frequency" (equivalent to half of the filter's cutoff frequency) was reduced from 8.19 Hz to 7.54 Hz. Frequency and ROCOF were calculated in accordance with the reference algorithms of the standard by using the first and second derivative of the measured phase, respectively. These derivatives are calculated using finite differences as shown in (D.3) and (D.4) of \cite{noauthor_ieeeiec_2018}.

\section{Hardware}
\begin{figure}[ht]
\centering
\includegraphics[width=3.49in]{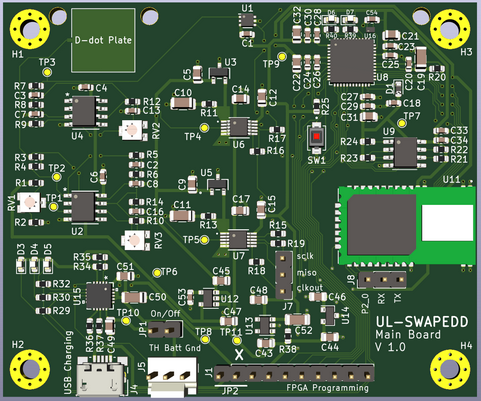}
\caption{Rendering of main PMU printed circuit board (PCB). The board measures 7.6 x 6.2 cm.}
\label{figPMU}
\end{figure}

The main board of the PMU is pictured in Fig. \ref{figPMU}. To support low-power component selection, it was designed on a custom printed circuit board (PCB) measuring 7.6 x 6.2 cm. A Lattice iCE40UP5k field-programmable gate array (FPGA) serves as the central processor and interface for the peripheral sensors and communications link. The following subsections detail other design decisions.

\subsection{Sensors}
A D-dot sensor similar to that in \cite{liu_wide-area-measurement_2016} measures the voltage from the electric fields underneath overhead power lines or near energized conductors without impacting the existing power system infrastructure in any way. The D-dot sensor measures the time-derivative of the electric flux density; this is proportional to the electric field generated by the voltage and capacitance of the power conductor(s). The sensor is composed of a 1-cm$^2$ metal plate connected to the input of a low-power, low input bias current OPA333 transconductance operational amplifier (op-amp). This metal plate is held at a virtual ground by the transconductance op-amp. When electric flux lines terminate on the metal plate, a charge is induced. The op-amp outputs a current ($i_{Ddot}$) that is proportional to the electric field according to
\begin{equation}
i_{Ddot}=\epsilon A \frac{d}{dt}E_\perp 
\end{equation}
where, $\epsilon$ is the permittivity of free space, $A$ is the area of the D-dot sensor plate, and $E_\perp$ is the electric field perpendicular to the D-dot sensor plate  (see (2.28) of \cite{chung_low_2017}). The transconductance op-amp drives this current through a large feedback resistor (20 M$\Omega$), generating a voltage signal that is proportional to the derivative of the conductor's electric field. For a sinusoidal electric field, the voltage is proportional to the electric field and frequency with a $\pi/2$ phase shift. The exact scaling from electric field to conductor voltage depends on the geometry of the conductor and the D-dot plate. 

An AH8501 Hall effect sensor measures conductor current. This sensor outputs an analog voltage with 8-bit resolution proportional to the magnetic flux density perpendicular to its face. Its sensitivity is 24.2 mT/V. The magnetic field is proportional to the current in the power conductor in accordance with the Biot-Savart law. The AH8501 Hall effect sensor can be enabled and disabled with the primary sample clock, making it a good candidate for low-power operation. The non-contact voltage and current sensors allow for fast, safe PMU installation without de-energizing the conductors.

A second-order Sallen-key low-pass filter (LPF) combined with a first-order RC filter creates a 60 dB/decade third-order filter with a cutoff frequency of 100 Hz to attenuate noise, harmonics, and interference from the output of the D-dot and Hall effect sensors. The TI ADS8866 16-bit analog-to-digital converter (ADC) was selected to sample the output of each filter due to its low power consumption and serial peripheral interface (SPI) that the iCE40UP5K FPGA natively supports. 

\subsection{Power Supply}
Charge management is performed by the Microchip MCP73781-2CCI/ML. This device is intended for a stand-alone system to provide load sharing and a lithium-ion/lithium-polymer battery charge management controller. A 3000-mAh, 3.7-V 18650 lithium-ion battery was selected for the onboard power source for the 1.2-V and 3.3-V low-dropout (LDO) voltage regulators. Future efforts will demonstrate the extended lifetime of the PMU with integrated energy harvesting.

\subsection{External Communication}
The PMU communicates with a host over Bluetooth Low Energy (BLE). The Microchip RN4870 BLE module includes an integrated antenna and simple universal asynchronous receiver-transmitter (UART) interface for data transmission between the FPGA and the RN4870. It also supports a low-power sleep state while duty cycling the communication link. 

This PMU can either transmit contiguous phasor data or function as an event recorder. In either case, the FPGA buffers 16-bit magnitude and phase data for the voltage and current channels in its 128-kB single-port random-access memory (SPRAM) which can hold nearly 5 minutes of data. When the buffer is full or an event is detected, the RN4870 wakes up and transmits the buffered data to a host within range of the BLE transmitter. When transmitting contiguous phasor data over the duty-cycled BLE link, the RN4870 is awake and transmitting with an 18\% duty cycle. When operating as an event recorder, the duty cycle (and power consumption) will decrease. Low duty cycle operation is suitable for gathering data for offline development of data-driven models. For applications requiring lower latency, the PMU can also increase the duty cycle to 100\% and transmit phasors as they are measured, albeit with a higher average power consumption.

\subsection{FPGA} \label{FPGA}
\begin{figure*}[b]
\centering
\includegraphics[width=7.1in]{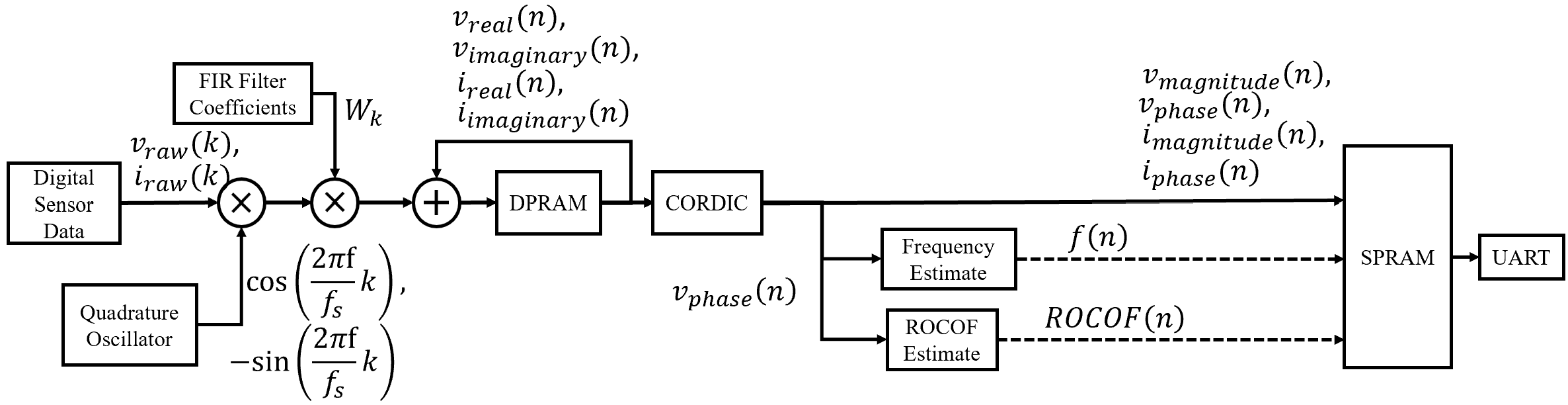}
\vspace{-2em}
\caption{FPGA architecture block diagram for the M-class phasor estimation algorithm. The sampled voltage and current data $(v_{\mathrm{raw}}, i_{\mathrm{raw}})$ are input at 3840 Hz (input sample index $k$). They are demodulated and multiplied by the corresponding filter coefficients ($W_k$) before accumulation into the real and imaginary voltage and current components ($v_{\mathrm{real}}, v_{\mathrm{imaginary}},i_{\mathrm{real}}, i_{\mathrm{imaginary}}$) (60-fps reporting rate phasor index $n$). The CORDIC converts the complex values from rectangular to polar form ($v_{\mathrm{magnitude}}, v_{\mathrm{phase}},i_{\mathrm{magnitude}}, i_{\mathrm{phase}}$), and then frequency ($\mathrm{f}(n)$) and ROCOF ($\mathrm{ROCOF}(n)$) are derived from the voltage phase.
}
\label{figFPGA_DSP}
\end{figure*}

Raw digital data collection and signal processing is performed on a Lattice Semiconductor iCE40UP5k FPGA. This is a small device, offering only 5280 lookup tables (LUTs), or anywhere from 5\% to 30\% of the logic of Xilinx and Intel FPGAs used by other PMU designs \cite{pinte_low_2015, romano_enhanced_2014, romano_high-performance_2017}. More importantly for the low-power design, this FPGA has a quiescent current of only 75 $\mu$A \cite{noauthor_ice40_nodate}.
Other useful features include eight 16-by-16-bit multipliers/32-bit adders used for data demodulation and filtering, 128 kB of SPRAM used for storing phasor values, and 15 kB of programmable dual-port RAM (DPRAM) used for storing filter coefficients and intermediate values \cite{noauthor_ice40_nodate}. The FPGA is favored over a low-power micro-controller due to the deterministic, cycle-accurate timing necessary for precise ADC sampling, deterministic throughput latency, and exceptionally low power. 

The reference M-class phasor estimation algorithm from Section \ref{Phasor Estimation Algorithms} was implemented and verified in Simulink. The equivalent Verilog code was generated from a block diagram using Simulink HDL Coder, a high-level synthesis (HLS) tool from Mathworks. Logic synthesis was performed using iCECube2 from Lattice Semiconductor. 

Fig. \ref{figFPGA_DSP} depicts the FPGA architecture. Due to the limited resources of the FPGA, the architecture reuses hardware extensively. To implement the quadrature oscillator, one full cycle of a 60-Hz cosine wave is stored in a DPRAM LUT, and the sine wave is derived from a $\pi/2$ offset from the corresponding cosine value. A single multiplier is time-multiplexed to demodulate both the voltage and current data. A second multiplier and filter coefficient DPRAM LUT implement the FIR filter. To reduce memory requirements, filter windows that overlap in time are interleaved, multiplying each demodulated value by one filter coefficient per overlapping filter window before the next sample is taken. This means only one sample needs to be stored in memory at a time rather than the entire window. This allows the architecture to scale efficiently with additional sensors and increased sample rate.

After filtering, the real and imaginary components are passed to a hardware-efficient coordinate rotation digital computer (CORDIC) algorithm that derives the magnitude and phase using only addition, subtraction, and bit shift operations \cite{andraka_survey_1998}. Finally, frequency and ROCOF are derived from the voltage signal's phase, and the magnitude and phase values of the phasor frame are stored in SPRAM for future transmission over BLE. The FPGA requires 3840 LUTs (73\%) to implement the M-class algorithm, inclusive of ADC control, phasor data buffering, UTC synchronization, and BLE interfacing. The remaining LUTs are available for higher-level processing of phasors (e.g., detecting power quality events).

\subsection{GPS UTC Synchronization}
A Ublox MAX-M8Q GPS receiver synchronizes the PMU sample clock to UTC. The GPS module outputs the date/time as well as a pulse per second (PPS) signal that has a rising edge precisely aligned with the start of every UTC second.  The MAX-M8Q supports a power save mode that allows for a low supply current at reduced update rates. To maintain synchronization, the PMU initiates one second of data collection on every rising edge of the PPS signal. To avoid jumps in the phase when intra-second timing error is corrected at the PPS edge (known as the ``time skew problem" \cite{zhang_synchrophasor_2014, qing_zhang_time_2012}), the true oscillator frequency is estimated by counting the number of 12-MHz clock cycles between consecutive PPS edges. A lookup table determines how to adjust the true sampling period using the measured clock error to minimize the timing error that accumulates between PPS edges. 

\subsection{Power Consumption}
Table \ref{power_breakdown} provides a coarse breakdown of the power consumed by each of the PMU's major components. The GPS consumed the most power of any active component (17.1 mW). Lower-power GPS modules or alternative methods of time synchronization will be necessary to reduce it further. The FPGA was the next leading contributor to active power. Reducing internal fixed-point precision, external oscillator frequency, or input/output (IO) pin supply voltage could further reduce the power. The sensors and ADCs consumed slightly less power than the FPGA, and the majority of this power is consumed by the AH8501 Hall effect sensor. BLE power will approach 0 mW if the PMU operates as an event recorder and events are infrequent. Power consumption from the voltage regulators may be reduced by using higher-efficiency switching regulators instead of LDOs, but the potential increase in noise on the analog components should be considered.

Table \ref{power_comparison} compares the power consumption of the proposed PMU with other PMUs developed in the literature. Some of these values had to be inferred from hardware datasheets because they were not reported explicitly. It can be seen from the table that the PMU presented in this paper uses an order of magnitude less power than any of the other solutions. As such, it can be powered through inexpensive energy harvesting with an 11-Wh battery and 300-mW solar panel costing under \$12. The size and weight of the proposed PMU are also small and low, respectively, making it considerably easier to transport and deploy than other PMUs. This allows for atypical use cases such as deploying it at one node for a time span of days or months before relocating to another node. 

\begin{table}[t]
  \begin{center}
    \caption{Power Consumption by Component}
    \vspace{-1.0em}
    \label{power_breakdown}
    \begin{tabular}{l|c}
      \hline
      \textbf{Component} & \textbf{Average Power} \\
      & \textbf{Consumption (mW)} \\
      \hline
      iCE40 FPGA & 4.6 \\
      Analog Sensors and ADCs & 3.4 \\
      Bluetooth Low Energy (BLE) & 1.6 \\
      Voltage Regulators, other Leakage & 4.1 \\
      MAX M8Q GPS &  17.1\textcolor{white}{a} \\
      \hline
      Total & 30.8\textcolor{white}{a} \\
      \hline
    \end{tabular}
  \end{center}
\end{table}

\begin{table}[b]
  \begin{center}
    \caption{Comparison with other single-phase PMUs}
    \vspace{-1.0em}
    \label{power_comparison}
    \begin{tabular}{c|c|c}
      \hline
      \textbf{Source} & \textbf{Measured Phasors} &  \textbf{Power Consumption (W)} \\
      \hline
      \cite{pinte_low_2015} & voltage & $>$2.600\textcolor{white}{II} \\
      \cite{thakre_economical_2020} & voltage &  $>$0.600\textcolor{white}{II} \\
      \cite{das_development_2016} & voltage and current &  0.396 \\
      This work & voltage and current &  0.031 \\
      \hline
    \end{tabular}
  \end{center}
\end{table}

\section{Hardware Performance Evaluation}
The PMU standard \cite{noauthor_ieeeiec_2018} outlines a variety of test cases to evaluate PMU performance. Steady-state tests measure performance under static conditions, including off-nominal fundamental frequency and under interference from harmonic or out-of-band signals. Dynamic tests evaluate performance during frequency ramps, magnitude/phase modulation, and magnitude/phase step changes.  The standard sets different maximum error limits depending on the specific test cases. PMU error is evaluated in terms of TVE, frequency error (FE), and ROCOF error (RFE). TVE quantifies combined error in magnitude and phase, and is reported in percentage. FE and RFE are the difference between the reported and true frequency and the reported and true ROCOF, and are measured in Hz and Hz/s, respectively. 

\begin{figure}[b]
\centering
\includegraphics[width=3.49in]{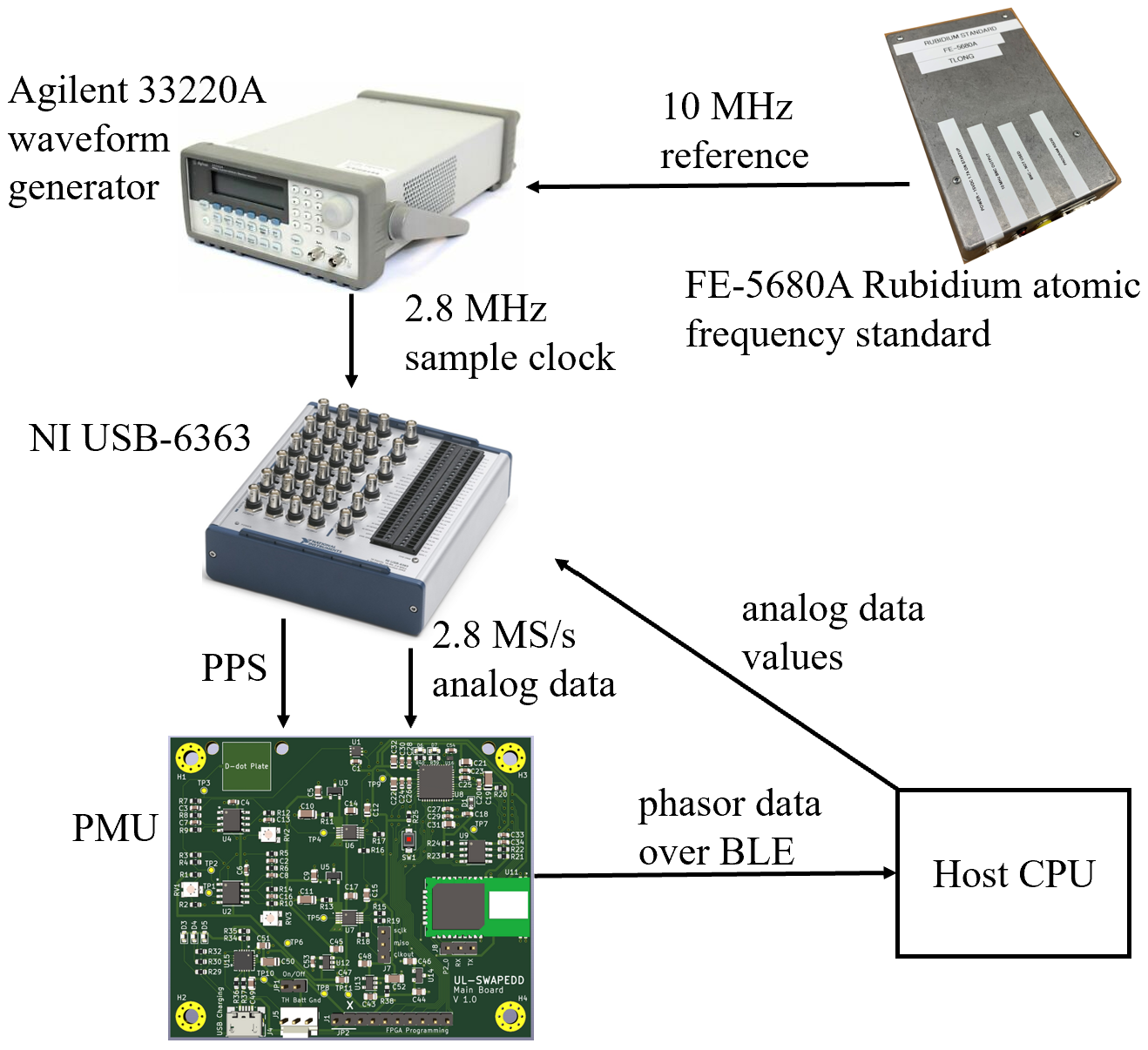}
\vspace{-2em}
\caption{Hardware-in-the-loop (HIL) test setup to evaluate PMU performance against the 2018 IEC/IEEE 60255-118-1 Standard. It generates low-voltage signals (0--3.3 V) with a total vector error (TVE) of 0.016\%.}
\label{figTestSetup}
\end{figure}

The hardware-in-the-loop (HIL) test setup, shown in Fig. \ref{figTestSetup}, was developed to evaluate the performance of the PMU. An FE-5680A Rubidium atomic frequency standard paired with an Agilent 33220A arbitrary waveform generator provided a precise output sample clock. An NI USB-6363 used the sample clock to produce the test signal with its 16-bit digital-to-analog converter (DAC). To simulate UTC-aligned signals, the analog test signals are generated in parallel with a synthetic GPS PPS signal to test PMU synchronization similar to the testbench in \cite{grando_synchrophasor_2018}. The host central processing unit (CPU) generated the sampled analog data values and ground truth, the former of which was sent to the NI USB-6363. The NI USB-6363 drove a low-voltage analog signal (0--3.3 V) at an output sampling rate of 2.8 mega-samples/s (MS/s) directly into the ADCs of the PMU. Finally, the measured phasors were transmitted over BLE to a host computer for evaluation. This test setup cannot generate voltages or currents large enough to be sensed by the PMU's analog D-dot or Hall effect sensors, so error contributed by the sensors was not analyzed. Future work will analyze analog sensor performance using the Fluke 6135A PMU calibration system available to the authors \cite{salls_statistical_2021-2}.

The PMU standard requires the ground truth of the input test signals be known with a TVE less than 0.1\% \cite{noauthor_ieeeiec_2018}. For test voltages between 0 V and 3.3 V, the NI USB-6363 introduces an analog gain error of 160 ppm, incurring a worst-case magnitude error of 0.016\%. Voltage offset error is ignored as the DC value will not impact the 60-Hz phasor magnitude. The analog output has a timing resolution of 10 ns, implying a phase error of $3.77 \mu$rad for a 60-Hz system. The FE-5680A frequency standard has a worst-case frequency error of 2 nHz, contributing negligible phase error. Combining magnitude and timing errors, the magnitude error dominates, and the TVE of the test setup is 0.016\%, which is well below the maximum allowable 0.1\% TVE.

A Python framework was developed to drive the HIL test setup. For each individual test, the test parameters and maximum error limits are determined by the PMU class desired for testing (M-class in our case). Where applicable, the reporting rate and nominal system frequency help determine the test parameters and maximum error limits as required by the PMU standard. Test data is generated using equations defined throughout Section 6 and Annex B.3 of \cite{noauthor_ieeeiec_2018}. Notably, testing of out-of-band interference follows the procedure defined in Section 6, Table 2 footnote. Additionally, testing of step changes in magnitude or phase follows the procedures defined in Sections 5.2.4, 5.2.5, and Annex C.4. The result of this testing setup is 523 individual M-class tests for evaluating compliance with the PMU standard requirements.

\section{Results}
\subsection{2018 IEC/IEEE 60255-118-1 Standard Compliance}

The PMU was evaluated according to the M-class tests specified in \cite{noauthor_ieeeiec_2018} for a 60-Hz system and a reporting rate of 60 fps. All tests were conducted with a 2-s lead-in time to ensure proper PPS resynchronization. Phasors were collected over a 5-s window after the lead-in time, with the exception of the 0.5-s step tests and 10-s frequency ramp tests. The test results are summarized in Table \ref{cool table}. Results in bold indicate error metrics that the PMU complied with.

In the steady-state off-nominal frequency tests, the PMU was tested over a bandwidth of 10 Hz with 0.1-Hz steps. The TVE remained below the error limits, but the maximum FE and RFE limits were exceeded. Frequency and ROCOF are calculated from the derivatives of the phase using finite differences, making them particularly sensitive to any noise in the system. Steady-state tests also have the lowest FE and RFE tolerance (0.005 Hz for frequency and 0.1 Hz/s for ROCOF), so they are the most difficult tests to pass in presence of noise.

During the harmonic distortion tests, the $2^{\mathrm{nd}}$ through $50^{\mathrm{th}}$ harmonics (120 Hz--3 kHz) were added individually to the nominal signal with 10\% of the nominal magnitude. As with the off-nominal frequency tests, TVE remained below the error limit. FE requirements were also met by the PMU. There is no RFE limit for the M-class harmonic distortion tests. 

Magnitude and phase modulation tests modulated the fundamental magnitude or phase with frequencies ranging from 0.1 Hz to 5 Hz in 0.1-Hz steps. All TVE, FE, and RFE requirements were satisfied, indicating that the PMU met the higher error limits of the modulation tests.

For the frequency ramp test, the input signal frequency was ramped from 55 Hz to 65 Hz at a rate of $1$ Hz/s. The PMU met TVE and FE requirements on the positive ramp but narrowly exceeded the FE limits on the negative ramp. The RFE requirement was not met in either of the ramp tests.

M-class specifications require a PMU to reject various out-of-band (OOB) interference signals under a range of nominal frequencies. OOB frequencies range from 10 Hz to 30 Hz and 90 Hz to 120 Hz. To evaluate performance, the input's fundamental frequency was varied from 57 Hz to 63 Hz for each interfering frequency. The maximum TVE, FE, and RFE were reported for each interfering frequency across all evaluated fundamental frequencies. TVE requirements were met, but only a subset of the 21 evaluated interfering frequencies met the FE requirements. There are no RFE requirements for these tests.

\begin{table}[t]
  \begin{center}
    \caption{PMU performance results}
    \vspace{-2.0em}
    \label{cool table}
    \includegraphics[width=3.5in]{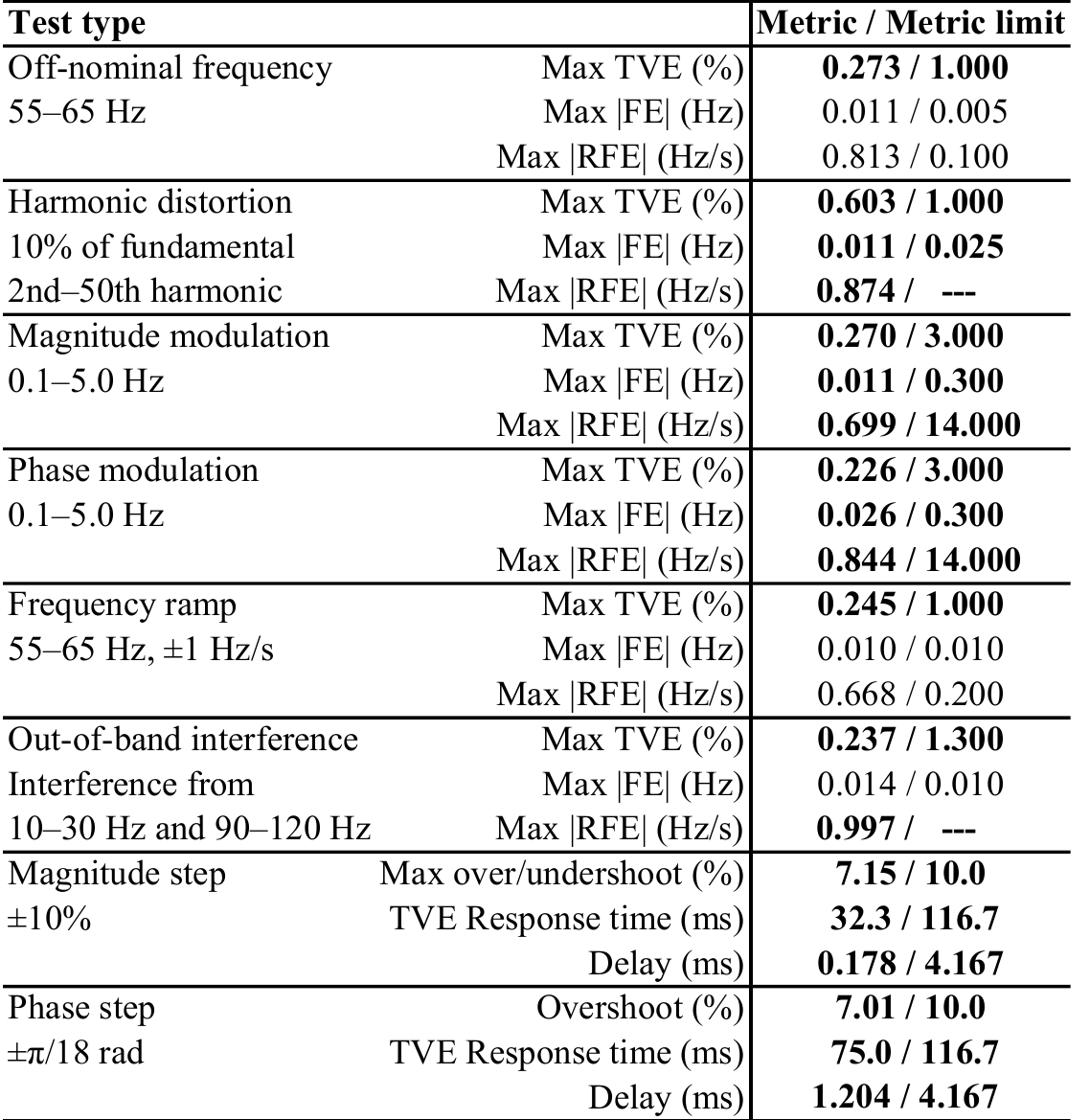}
  \end{center}
\end{table}

Magnitude and phase step tests are evaluated from the overall step response. This was obtained by first recording the step response from a step occurring at 20 even timesteps between consecutive reporting intervals. The step times were then aligned, allowing for a high-resolution view of the step response as described in Annex C.4 of \cite{noauthor_ieeeiec_2018}. Step tests set limits on the overshoot/undershoot (the maximum allowable over/underestimate of the stepped parameter), response time (time that TVE, FE, and RFE exceed steady-state limits during the transient), and delay time (time between step and measurement of a 50\% step transition). The magnitude step is 10\% of nominal, and the phase step is $\pi$/18 rad.
The M-class overshoot limit was satisfied, and the delay time was less than 1 ms for both the magnitude and phase step, staying below the 4.2-ms delay limit. The PMU has a TVE response time of 32.3 ms for the magnitude step and 75.0 ms for phase, remaining below the 116.7-ms limit. The PMU is unable to meet the steady-state FE and RFE requirements at nominal frequency, so response times could not be evaluated for these metrics. Similar results were obtained for the negative step tests. 

The PMU successfully met the TVE requirements across all tests. It did not meet the FE and RFE requirements unless they were relaxed from the steady-state requirements, as in the modulation tests. Additional hardware evaluation indicated that the ADCs have only 8 effective number of bits (ENOB) of the ideal 16. A redesign of the analog front-end external ADC reference voltage circuitry is expected to improve the ADC performance, further reducing the TVE, FE, and RFE with a negligible change in power consumption.

\subsection{Measurement of Real Load}

\begin{figure}[h!]
\centering
\includegraphics[width=3.49in]{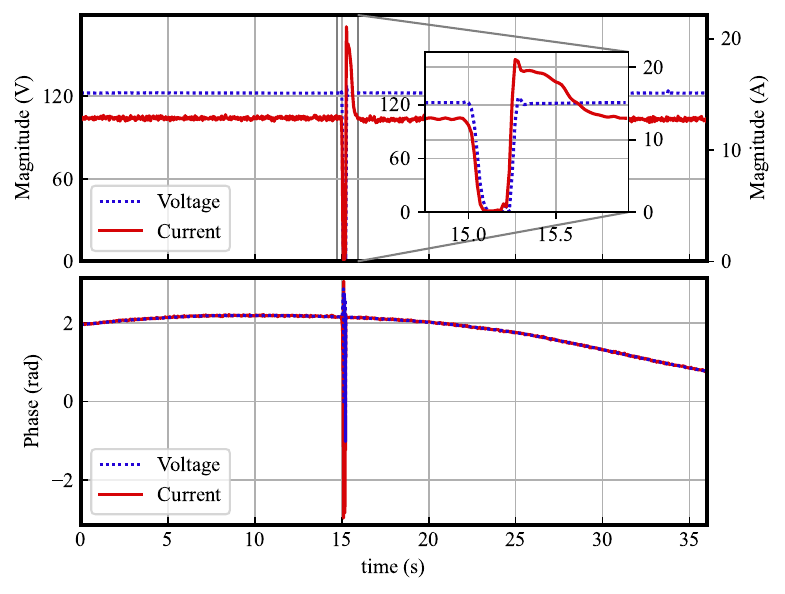}
\vspace{-2em}
\caption{PMU output from the supply cable of a 1500-W space heater. Voltage (dotted blue) and current (solid red) magnitude and phase were recorded. A fault was simulated by momentarily switching the voltage
supply off around $t=15$ s.}
\label{heater phasors}
\end{figure}

The proposed PMU's measurement of an actual load is shown in Fig. \ref{heater phasors}. This figure depicts the PMU measurements from the single-phase supply cable of a 1500-W space heater. To demonstrate dynamic measurements, a nearby fault was simulated by rapidly toggling the voltage input to the space heater. The PMU clearly captures the transient behavior of the voltage drop and subsequent spike in current after power is restored. Such an event could be stored on the FPGA's SPRAM for future transmission or transmitted immediately to the host via BLE.

\section{Conclusion}
In contrast to previous works that explored PMU design trade-offs between cost and performance, this work is the first to prioritize accuracy and low power consumption for battery-powered operation. The PMU's maximum TVE was on the order of 0.25\% for most tests, and hardware improvements will further increase the accuracy.
The proposed design reduced power consumption by more than an order of magnitude compared to other PMUs in the literature. These optimizations make long-term PMU deployment away from reliable power supplies a reality.

Future work will explore alternative phasor estimation algorithms, lower-power UTC synchronization options, sensor calibration, and distribution system applications of the PMU measurements. The low-power design will enable ubiquitous PMU placement in areas where it is not presently feasible.

 \section*{Acknowledgment}
 The authors would like to thank Prof. Yang Weng from Arizona State University and David Hull from the DEVCOM Army Research Laboratory for their valuable discussions.

\bibliographystyle{IEEEtran}
\bibliography{IEEEabrv,references}

\begin{thebibliography}{10}
\providecommand{\url}[1]{#1}
\csname url@samestyle\endcsname
\providecommand{\newblock}{\relax}
\providecommand{\bibinfo}[2]{#2}
\providecommand{\BIBentrySTDinterwordspacing}{\spaceskip=0pt\relax}
\providecommand{\BIBentryALTinterwordstretchfactor}{4}
\providecommand{\BIBentryALTinterwordspacing}{\spaceskip=\fontdimen2\font plus
\BIBentryALTinterwordstretchfactor\fontdimen3\font minus \fontdimen4\font\relax}
\providecommand{\BIBforeignlanguage}[2]{{%
\expandafter\ifx\csname l@#1\endcsname\relax
\typeout{** WARNING: IEEEtran.bst: No hyphenation pattern has been}%
\typeout{** loaded for the language `#1'. Using the pattern for}%
\typeout{** the default language instead.}%
\else
\language=\csname l@#1\endcsname
\fi
#2}}
\providecommand{\BIBdecl}{\relax}
\BIBdecl

\bibitem{follum_phasors_2021}
J.~Follum, L.~Miller, P.~Etingov, H.~Kirkham, A.~Riepnieks, X.~Fan, and E.~Ellwein, ``Phasors or {{Waveforms}}: {{Considerations}} for {{Choosing Measurements}} to {{Match Your Application}},'' Apr. 2021.

\bibitem{phadke_synchronized_2017}
A.~G. Phadke and J.~S. Thorp, \emph{Synchronized {{Phasor Measurements}} and {{Their Applications}}}.\hskip 1em plus 0.5em minus 0.4em\relax {Springer}, 2017.

\bibitem{kezunovic_application_2014}
M.~Kezunovic, S.~Meliopoulos, V.~Venkatasubramanian, and V.~Vittal, \emph{Application of {{Time-Synchronized Measurements}} in {{Power System Transmission Networks}}}.\hskip 1em plus 0.5em minus 0.4em\relax {Springer}, 2014.

\bibitem{thakre_economical_2020}
M.~P. Thakre, P.~S. Jagtap, and M.~Sharma, ``Economical synchrophasor data acquisition system for {{WAMS}} implementations,'' in \emph{2020 {{International Conference}} on {{Power}}, {{Energy}}, {{Control}} and {{Transmission Systems}} ({{ICPECTS}})}, Dec. 2020, pp. 1--6.

\bibitem{romano_high-performance_2017}
P.~Romano, M.~Paolone, T.~Chau, B.~Jeppesen, and E.~Ahmed, ``A high-performance, low-cost {{PMU}} prototype for distribution networks based on {{FPGA}},'' in \emph{2017 {{IEEE Manchester PowerTech}}}.\hskip 1em plus 0.5em minus 0.4em\relax {Manchester, United Kingdom}: {IEEE}, Jun. 2017, pp. 1--6.

\bibitem{liu_wide-area-measurement_2016}
Y.~Liu, L.~Zhan, Y.~Zhang, P.~N. Markham, D.~Zhou, J.~Guo, Y.~Lei, G.~Kou, W.~Yao, J.~Chai, and Y.~Liu, ``Wide-area-measurement system development at the distribution level: {{An FNET}}/{{Grideye}} example,'' \emph{IEEE Transactions on Power Delivery}, vol.~31, no.~2, pp. 721--731, Apr. 2016.

\bibitem{das_development_2016}
H.~P. Das and A.~K. Pradhan, ``Development of a micro-phasor measurement unit for distribution system applications,'' in \emph{2016 {{National Power Systems Conference}} ({{NPSC}})}.\hskip 1em plus 0.5em minus 0.4em\relax {Bhubaneswar, India}: {IEEE}, Dec. 2016, pp. 1--5.

\bibitem{patterson_inductively_2021}
J.~Patterson and A.~Pal, ``An {{Inductively Powered Line-Mounted Time-Synchronized Micro Point-on-Wave Recorder}},'' in \emph{2021 {{IEEE Power}} \& {{Energy Society General Meeting}} ({{PESGM}})}, Jul. 2021, pp. 1--5.

\bibitem{noauthor_ieeeiec_2018}
``{{IEEE}}/{{IEC International Standard-Measuring}} relays and protection equipment {\textendash} {{Part}} 118-1: {{Synchrophasor}} for power systems {\textendash} {{Measurements}},'' 2018.

\bibitem{chung_low_2017}
H.~E. Chung, ``Low frequency electric field imaging,'' Ph.D. dissertation, Arizona State University, Jul. 2017.

\bibitem{pinte_low_2015}
B.~Pinte, M.~Quinlan, and K.~Reinhard, ``Low voltage micro-phasor measurement unit ({{$\mu$PMU}}),'' in \emph{2015 {{IEEE Power}} and {{Energy Conference}} at {{Illinois}} ({{PECI}})}.\hskip 1em plus 0.5em minus 0.4em\relax {Champaign, IL, USA}: {IEEE}, Feb. 2015, pp. 1--4.

\bibitem{romano_enhanced_2014}
P.~Romano and M.~Paolone, ``Enhanced interpolated-{{DFT}} for synchrophasor estimation in {{FPGAs}}: {{Theory}}, implementation, and validation of a {{PMU}} prototype,'' \emph{IEEE Transactions on Instrumentation and Measurement}, vol.~63, no.~12, pp. 2824--2836, Dec. 2014.

\bibitem{noauthor_ice40_nodate}
``{{iCE40 UltraPlus}} - {{Lattice Semiconductor}},'' https://www.latticesemi.com/products/fpgaandcpld/ice40ultraplus.

\bibitem{andraka_survey_1998}
R.~Andraka, ``A survey of {{CORDIC}} algorithms for {{FPGA}} based computers,'' in \emph{Proceedings of the 1998 {{ACM}}/{{SIGDA}} Sixth International Symposium on {{Field}} Programmable Gate Arrays - {{FPGA}} '98}.\hskip 1em plus 0.5em minus 0.4em\relax {Monterey, California, United States}: {ACM Press}, 1998, pp. 191--200.

\bibitem{zhang_synchrophasor_2014}
Q.~F. Zhang and V.~M. Venkatasubramanian, ``Synchrophasor time skew: {{Formulation}}, detection and correction,'' in \emph{2014 {{North American Power Symposium}} ({{NAPS}})}.\hskip 1em plus 0.5em minus 0.4em\relax {Pullman, WA, USA}: {IEEE}, Sep. 2014, pp. 1--6.

\bibitem{qing_zhang_time_2012}
{Qing Zhang}, V.~Vittal, G.~Heydt, Y.~Chakhchoukh, N.~Logic, and S.~Sturgill, ``The time skew problem in {{PMU}} measurements,'' in \emph{2012 {{IEEE Power}} and {{Energy Society General Meeting}}}.\hskip 1em plus 0.5em minus 0.4em\relax {San Diego, CA}: {IEEE}, Jul. 2012, pp. 1--6.

\bibitem{grando_synchrophasor_2018}
F.~L. Grando, A.~E. Lazzaretti, G.~W. Denardin, M.~Moreto, and H.~V. Neto, ``A {{Synchrophasor Test Platform}} for {{Development}} and {{Assessment}} of {{Phasor Measurement Units}},'' \emph{IEEE Transactions on Industry Applications}, vol.~54, no.~4, pp. 3122--3131, Jul. 2018.

\bibitem{salls_statistical_2021-2}
D.~Salls, J.~R. Torres, {Antos}, C.~Varghese, J.~Patterson, and A.~Pal, ``Statistical {{Characterization}} of {{Random Errors Present}} in {{Synchrophasor Measurements}},'' in \emph{2021 {{IEEE Power}} \& {{Energy Society General Meeting}} ({{PESGM}})}.\hskip 1em plus 0.5em minus 0.4em\relax {Washington, DC, USA}: {IEEE}, Jul. 2021, pp. 01--05.

\end{thebibliography}

\end{document}